\documentstyle[12pt]{article}
\renewcommand{\theequation}{\arabic{section}.\arabic{equation}}
\newcommand{\be}{\begin{equation}}
\newcommand{\ee}{\end{equation}}
\newcommand{\YBE}{Yang-Baxter equations}
\newcommand{\BA}{Bethe ansatz}

\newcommand{\CFT}{conformal field theory}
\newcommand{\NP}{Nucl. Phys.~}
\newcommand{\CMP}{Commun. Math. Phys.~}
\newcommand{\LMP}{Lett. Math. Phys.~}
\newcommand{\IJMP}{Int. Journ. Mod. Phys.~}

\newcommand{\Sm}{{\sl S}-matrix}
\newcommand{\Sms}{{\sl S}-matrices}
\newcommand{\PEST}{purely elastic scatterring theory}
\newcommand{\IM}{integral of motion}
\newcommand{\IMs}{integrals of motion}
\newcommand{\bea}{\begin{eqnarray}}
\newcommand{\eea}{\end{eqnarray}}
\newtheorem{T}{Theorem}
\newtheorem{rem}{Remark}
\newtheorem{prop}{Proposition}

\newtheorem{col}{Corollary}
\newtheorem{con}{Conjecture}
\newcommand{\e}{\hat{e}}

\newcommand{\f}{\hat{f}}

\newcommand{\h}{\hat{h}}
\newcommand{\K}{\hat{K}}
\newcommand{\hh}{\frac{\h}{2}}
\newcommand{\half}{\frac{1}{2}}
\newcommand{\imat}{\dot{\imath}}
\newcommand{\g}{\gamma}
\newcommand{\G}{\Gamma}
\newcommand{\pauli}{\hat{\sigma}}
\newcommand{\qsig}[1]{e^{\dot{\imath} \frac{\gamma}{2}
\pauli^3_{#1}}}
\newcommand{\qsigm}[1]{e^{-\dot{\imath}\frac{\gamma}{2}
\pauli^3_{#1} }}
\newcommand{\et}[2]{e^{{#1} \nt_{#2} }}
\newcommand{\qh}[1]{e^{\dot{\imath}
\frac{\gamma}{2} \h_{#1} }}
\newcommand{\qhm}[1]{e^{-\dot{\imath}
\frac{\gamma}{2} \h_{#1} }}
\newcommand{\Delb}{\overline{\Delta}}
\newcommand{\delb}{\overline{\delta}}
\newcommand{\exh}[1]{\exp{({#1}\imat
\frac{\g}{2}\Delta_{12}(\h))}}
\newcommand{\Pro}{{\cal P}^{~J}_{J_1J_2} }
\newcommand{\pro}{{\cal P}^{~L}_{J_1J_2} }
\newcommand{\Ppro}{|J_2J_1\rangle {\sf P}^{J}\langle J_1J_2|}
\newcommand{\U}{{\cal U}}
\newcommand{\th}{\theta}
\newcommand{\al}{\alpha}
\newcommand{\nt}{\nu \theta}

\newcommand{\gn}{\frac{\g}{\nu} }
\newcommand{\ip}{\imat \pi}
\newcommand{\ig}{\imat \g}
\newcommand{\A}{\overline{a}}
\newcommand{\B}{\overline{b}}
\newcommand{\J}{\overline{J}}

\begin{document}
\begin{titlepage}

\title{{\Large
Quantum Group Analysis of the Bound States in the Strong Coupling
Regime of the Modified Sine-Gordon Model}\thanks{PACS:
05.50.+q,02.90.+p,75.10 Jm}}

\author{{\large Sergei V. Pokrovsky}\thanks{E--mail
pokrovsky@binah.cc.brandeis.edu}\\
\smallskip
{\normalsize
{\it Physics Department, Brandeis University,}}\\
{\normalsize{\it Waltham, MA 02254-9110 U.S.A.}}}

\date{}
\maketitle

\begin{abstract}
A quantum group analysis is applied to the Sine-Gordon model (or may
be its version) in a strong-coupling regime. Infinitely many bound
states are found together with the corresponding \Sms. These new
solutions of the \YBE~are related to some reducible representations of
the quantum $sl(2)$ algebra resembling the Kac-Moody algebra
representations in the Wess-Zumino-Witten-Novikov \CFT. \end{abstract}

\end{titlepage}
\addtocounter{section}{-1}
\section[in]{Introduction.}

A concept of the exact integrabilty played an outstanding role both in
statistical mechanics and in the quantum field theory in two
dimensions. Various aspects of it have become recently an object of a
new close examination due to the observation by A.~Zamolodchikov
\cite{Z} that certain deformations of conformal field theories may
posses infinitely many non-trivial conserved charges giving rise to
integrable field theories, whose \Sms ~factorize and satisfy the
\YBE~(YBE).

Thus studying some deformations of the Virasoro algebra and the higher
ones might lead to the  essentially new understanding of the universal
behavior in the vicinity of criticality. Knowledge of different
near-critical regimes is essential for solving a problem of
classification of the critical points being attacked since the advent
of conformal invariance \cite{BPZ}.

One of the first \PEST ~investigated in detail was a sine-Gordon
model. It had received a lot of attention as a favorite toy of the
theorists in the late seventies (see \cite{ZZ} and references
therein).  Nevertheless, it is shown in the present paper that in a
strong-coupling regime this model (or may be its version) exhibits a
rich structure of the infinitely many bound states almost
unexplored.\footnote{The existence of these bound states was mentioned
in \cite{KiR}.}

The structure of these bound states resembles that of the Kac-Moody
algebra representations in the Wess-Zumino-Novikov-Witten \CFT.
Moreover, the similarity  to the WZNW \CFT ~is by no means accidental
since the non-local conserved currents of the sine-Gordon model
derived in \cite{ResSm,BerCl,Mathur} generate the $q$-deformed loop
algebra which is isomorphic to the Kac-Moody one up to the central
element.

However, the absence of central extension leads to the essential
distinctions as far as the structure of the representations is
concerned. The bound states discussed below are associated with the
irreducible and some special reducible representations of the quantum
$sl(2)$ algebra rather than with the integrable highest weight modules
of the affine $A^{(1)}_1$ algebra. For some of these super-multiplets
the new solutions of the YBE are found. It is shown that the number of
these solutions is in fact infinite and finding the higher ones
through the more elementary recursively one encounters more and more
reducible representations.

Such reducible representations have been known in the \PEST
{}~al\-ready. For the ratio\-nal \Sms~with the higher symmetries they
were dis\-cussed by Karowsky \cite{Kar} and in a series of works
\cite{OW,RW,ORW} dealing with the classification of the integrable
vertex models in terms of \BA. A general form of the \BA~entirely in
terms of the representation theory of the simple Lie algebras has been
conjectured in these papers. The limitations of the proposed formulas
and the underlying algebraic structures remained somewhat obscure. I
hope that the approach used in the present paper will help to
elucidate the subject to some extent.

The technique used is a quantum group analysis introduced by Kulish
and Reshetikhin just in the context of the higher spin representations
of the sine-Gordon model \cite{KR} and developed in a series of works
by Jimbo \cite{J1,J2,J3} who first realized the relation of the
algebra considered to the affine algebra. The innovations I propose
below make it possible in more or less regular way to produce new
solutions of the \YBE ~recursively.

Despite the formal absence of the anomaly in the current algebra there
is a strong evidence that it manifests itself indirectly since the
requirements of unitarity and crossing permit only a finite number of
the irreducible multiplets while the reducible ones may be regarded in
a sense as their descendants. This conjecture of the dynamically
generated anomaly is discussed in the end of the paper.

\medskip

The paper is organized as follows. In the next section a brief review
of the most important \PEST~concepts is given in order to make the
paper self-consistent and to fix the convenient notations for what it
follows. The loop $sl(2)$ quantum algebra is derived from the YBE and
discussed thoroughly in the sect. 3. Using these results the fusion
procedure for the \Sms~and the massive fusion rules are considered in
the sect. 4. The unsolved problems and the prospects for the future
investigations are summarized in the last section. A cumbersome proof
of the fusion theorem is presented in the Appendix.

\setcounter{section}{0}
\section[gen]{Generalities of the \PEST.}

Some general facts  concerning the \PEST~are collected in this section
in order to fix notations mainly. First, a conventional
parametrization for the 2D relativistic systems of the energy and
momentum through the rapidity is used
\be
E = m \cosh{\th}; ~~~~~~~~~ P = m \sinh{\th},
\ee
where $m$ is the mass of the particle.

\subsection[kin]{Bootstrap kinematics.}

Consider a collision of two particles, say "$a$" and "$b$", with the
masses $m_a$ and $m_b$ respectively, in the rest frame of their mass
center. Suppose that a third particle "$c$" with the mass $m_c$ is
produced as an intermediate bound state in course of this scattering
process. The requirement of the energy-momentum conservation leads to
the following equations
\be
m_c = m_a \cosh{\theta_a} + m_b \cosh{\theta_b};
{}~~~~~~~~~{}
0= m_a \sinh{\theta_a} + m_b \sinh{\theta_b}.
\label{cons:e-m}
\ee
These constraints determine the rapidities of colliding particles
$\theta_a$ and $\theta_b$ in terms of masses $m_a$, $m_b$ and $m_c$.
It is more convenient, however, to parametrize the masses themselves
in terms of rapidities.
\be
m_a = -m_0 \sinh{\theta_b};~~~~~~~
m_b = m_0 \sinh{\theta_a}; ~~~~~~~
m_c = m_0 \sinh{(\theta_a - \theta_c)}.
\label{mass-rap}
\ee
Once the particle "$c$" is a bound state its mass should be less then
the total mass of its parents. Hence,
\[\frac{m_a+m_b}{m_c} = \frac{\cosh {\frac{\theta_a - \theta_b}{2}}}{
\cosh {\frac{\theta_a + \theta_b}{2}}}>1 \]
which becomes possible only when the rapidities $\theta_a$ and
$\theta_b$ take the purely imaginary values. Put
\be
\theta_a = \dot{\imath} U^{\B}_{c\A}; ~~~~~~~~~
\theta_b = -\dot{\imath} U^{\A}_{\B c}.
\label{Uabc}
\ee
Then
\be
m_a = m \sin{U^{\A}_{\B c}}; ~~~~~~~{}
m_b = m \sin{U^{\B}_{c\A}}; ~~~~~~~{}
m_c = m \sin{U^c_{ab}},
\label{mass-par}
\ee
\be
U^c_{ab} = U^{\A}_{\B c} + U^{\B}_{c\A}.
\label{add}
\ee
Here the bar over the indices means charge conjugation. This
parametrization may be interpreted geometrically in a very transparent
way. Three Euclidean two-vectors \(P_c = P_a + P_b \) form a triangle.
Hence, for any of them the ratio of its length to the sine of the
angle between two remaining ones is the same.

The intermediate virtual bound state reveals itself as a pole of a
two-particle \Sm~
\be
S_{a,b}(\theta) = \frac{(f_{ab}^c)^2}{\theta - \imat U^c_{ab} }|c \rangle
\langle c|  +
\cdots \; .
\label{pole}
\ee
Consider a three-particle amplitude
\be
S_{d,a,b}(\theta_d, \theta_a, \theta_b) = S_{a,b}(\theta_{da})
S_{a,b}(\theta_{ab}) S_{a,b}(\theta_{db}) ,
\label{3part}\ee
where
\(  \theta_{da} = \theta_d - \theta_a,~etc. \)
Examining its dependence on the argument $\theta_{ab}$ one finds a
pole at \(\theta_{ab} = \dot{\imath} U^c_{ab} \). In the vicinity of
this pole the amplitude is dominated by the scattering through the
bound state $c$-channel. Therefore the residue of the above pole is
considered as a scattering amplitude of $d$ and $c$ particles.
However, the value of the rapidity $\theta_c$ is not fixed by the
above resonance condition and is determined via the kinematical
relations (\ref{Uabc})
\be
\theta_{ac} = - \dot{\imath} U^{\B}_{c\A};  ~~~~~~~{}
\theta_{bc} = \dot{\imath} U^{\A}_{\B c}.
\label{resonance}
\ee
leading to the following  celebrity bootstrap equation \cite{Z}
\be
S_{d,c}(\theta) = P^c_{ab}\;
S_{a,c}(\theta - \dot{\imath} U^{\B}_{c\A})
S_{b,c}(\theta + \dot{\imath} U^{\A}_{\B c})\; P^c_{ab}
\label{boot}
\ee
Here a projection operator $P^c_{ab}$ on the state
$|c \rangle $ is introduced.

\subsection[spectr]{The \IMs~and the mass spectrum.}

The imaginary rapidities (or Euclidean angles) $U^a_{bc}$ etc.
extracted from the solution of the bootstrap equations (\ref{boot})
define not only the mass spectrum but also the spectra of the other
\IMs ~\cite{Z,ZZ}. The existence of an infinite number of conservation
laws is absolutely necessary in order for the \Sm~to be factorizable.
The commuting \IMs~ are usually supposed to be local, additive and
Lorentz-covariant. Therefore the spin-$s$ \IM~${\cal I}_s$ acts on
some $N$-particle in-state $|A_1(\theta_1)A_2(\theta_2) \cdots
A_N(\theta_N)\rangle$ just as if it acted on each of the one-particle
states $|A_n(\theta_n) \rangle$ separately:
\be
{\cal I}_s |A_1(\theta_1) \cdots A_N(\theta_N) \rangle =
\sum_{n=1}^{N} I_{n}^s \exp{(s \theta_{n})}
|A_1(\theta_1)\cdots A_N(\theta_N) \rangle
\label{IM}
\ee
The exponential dependence of the one-particle eigenvalues on
corresponding rapidities is dictated by the above mentioned
Lorentz-covariance. The pre-exponential factors $I_{n}^s$ cannot be
determined kinematically. However, they are related to each other in
the same manner as the masses of the corresponding particles. Really,
consider again the process of two-particle scattering through the
bound state discussed in the beginning of the previous subsection. In
conformity with the eq.~(\ref{IM}) the value of the \IM~${\cal I}_s$
in the initial state \(|A_a(\theta_a)A_b(\theta_b)\rangle\) is given
by
\[I_{a}^s \exp{(s \theta_{a})} + I_{b}^s \exp{(s \theta_{b})}. \]
In the vicinity of the resonance~(\ref{pole})
its value in the final state
\[ S_{a,b}(\theta_{ab}) |A_a(\theta_a)A_b(\theta_b)\rangle \]
is dominated by the pole contribution with the residue
\[I_{a}^s \exp{(s \theta_{a})} \]
coming from the bound state channel. Hence, by virtue of
the condition~(\ref{resonance})
\be
I_{c}^s =
I_{b}^s \exp{(-\imat s U^{\B}_{c\A})}
+ I_{a}^s \exp{(\imat s U^{\A}_{\B c})}
\label{Uconstr}
\ee
This constraint can be resolved as follows:
\be
I_a^s = I \sin{(sU^{\A}_{\B c})}; ~~~~~~~{}
I_b^s = I \sin{(sU^{\B}_{c\A})}; ~~~~~~~{}
I_c^s = I \sin{(sU^c_{ab})}.
\label{IM-par}
\ee
A parametrization of masses~(\ref{mass-par}) is certainly a particular
case of~(\ref{IM-par}) for $s=\pm 1$. Since in general a given sort of
particles, e.g. "$c$", may participate in several types of reactions
either as a real particle or an intermediate bound state its mass and
any other \IM ~may be represented as a product over all such
three-particle virtual processes:
\be
m_c = \prod_{a,b} m_0 \sin{U^c_{ab}}; ~~~~~~~~~~~{}
I_c^s = \prod_{a,b} I_0 \sin{(sU^c_{ab})}
\label{gen.IM-par}
\ee
The angles $U^c_{ab}$ are usually rational multiples of $\pi$ if a
number of particles is finite in a theory. For this reason the \IM
{}~${\cal I}_s$ with the spin $s$ divisible by all denominators of these
rational multipliers vanish identically for any multi-particle
asymptotic state. This fact served as a foundation for a work \cite{Z}
where a connection of a certain \PEST~to the critical Ising model
perturbed by a magnetic field was established.

\subsection[fact,unit,cross]{The factorizability the unitarity
and the crossing.}

When the \Sm ~cannot be diagonalized and thus cannot be reduced to
some pure phase factors to a number of new restrictions arise.

Consider a scattering of two particles $\alpha$ and $\beta$. Let
$\alpha$ belongs to the multiplet $a$ consisting of $n_a$ particles
with the equal masses $m_a$ and $\beta$ belongs to the multiplet $b$
consisting of $n_b$ particles having mass  $m_b$. The different
particles $\alpha$ and $\beta$ are labeled by the isotopic quantum
numbers including besides the common multiplet indices $a$ and $b$
also internal indices $M_{\alpha}$ and $M_{\beta}$ allowing to
distinguish the particles inside each of the multiplets and referred
below as the isospin projections. The latter may be changed after
scattering. Therefore an \Sm~$S_{[a,b]}(\th _a-\th_b)$ may be viewed
as a mapping: ${\sf V}_a \otimes {\sf V}_b \rightarrow {\sf V}_a
\otimes {\sf V}_b$ where ${\sf V}_a$ and ${\sf V}_b$ are the
$n_a$-dimensional and $n_b$-dimensional spaces of one-particle
in-states.

A factorization hypothesis \cite{ZZ} for the multi-particle scattering
in the completely integrable system implies that the net $N$-particle
amplitudes do not depend on the order in which the intermediate
two-particle processes take place. This assumption specified for a
$3$-particle scattering imposes stringent constraints on the
admissible two-particle \Sms~known as the \YBE~(YBE) \cite{ZZ,Bax}.
\bea
S_{[a,b]}(\th_a-\th_b) S_{[a,c]}(\th_a-\th_c) S_{[b,c]}(\th_b-\th_c) =
{}~~~~~~~~~~~~~~~~~~~~~~~~~~~~~~~\nonumber\\
{}~~~~~~~~~~~~~~~~~~~~~~~~~~~~~~~ 	{}
S_{[b,c]}(\th_b-\th _c) S_{[a,c]}(\th_a-\th_c) S_{[a,b]}(\th_a-\th_b).
\label{ybe}
\eea
The YBE combined with a standard initial condition
\be
S_{[a,b]}(\th)|_{\th= 0} = {\sf P}_{[a,b]},
\label{init}
\ee
where ${\sf P}_{[a,b]}$ is the permutation operator for two identical
multiplets $a$ and $b$, lead to the following relations for their
solutions
\be
S_{[b,a]}(-\th)S_{[a,b]}(\th) =  f(\th) {\hat I}
\label{unit}
\ee
where ${\hat I}$ is an identity operator. This condition is almost
equivalent to the unitarity provided the real analyticity
\be
S_{[b,a]}(-\th^{\ast}) = S_{[a,b]}^{\dagger}(\th)
\label{analyt}
\ee
is respected.

One more conventional dynamical principle in the \PEST~is that of the
crossing symmetry. We shall also use it in a slightly generalized form
\be
S_{[a,b]}(\th) =  U^{-1} S_{[\overline{b},a]}(\ip-\th) U
\label{cross}
\ee
where a $b$-antiparticle is denoted by $\overline{b}$ and $U$ is
the matrix of some unitary transformation.
\section[Aff]{Affine ${\cal U}_q( sl(2))$ symmetry  of the YBE.}
\setcounter{equation}{0}
\subsection[SG]{The sine-Gordon S-matrix.}

The simplest trigonometric solution of the YBE for the case when all
three particles belong to the same doublet (i.e. all the spaces ${\sf
V}_a$, ${\sf V}_b$ and ${\sf V}_c$ are $2$-dimensional) has been known
since long ago \cite{Bax}.
\bea
S_{~[a,b]}^{(\half,\half)}(\th) = \qquad \qquad \qquad \qquad
\qquad \qquad \qquad \qquad \qquad \qquad \qquad \qquad
\nonumber\\
 Z(\th) \left\{
 \sinh {\left( \nt  +  \imat \frac{\gamma}{2} (1 +
	 \pauli ^{3} _a \pauli ^{3}_b) \right) }
+ \imat \sin{\gamma} \left(e^{\nt} \pauli ^{-} _a
	\pauli^{+}_b
+  e^{- \nt} \pauli ^{+}_a \pauli ^{-}_b \right) \right\},
\label{sixvert}
\eea
where both $\pauli^{\lambda}_a, \pauli^{\lambda}_b (\lambda = \pm,3)$
map $sl(2) \rightarrow End({\sf V}_a \otimes {\sf V}_b) $: \[ \pauli_a
= \pauli \otimes {\hat I};\;\;\; \pauli_b = {\hat I} \otimes \pauli \]
and all the operators $\pauli^{3}, \pauli^{\pm }$ are just the Pauli
matrices. The overall scalar factor $Z(\th)$ doesn't affect the YBE
but is essential for the unitarity of the \Sm~  and its appropriate
behavior in the ultra-relativistic limit. In particular it serves to
eliminate the function $f(\th)$ in the eq.~(\ref{unit}). Notice that
the above \Sm~differs from the conventional one of the six-vertex
model \cite{Bax} but may be related to the latter via a simple
transformation:
\be
S_{[a,b]}^{(\half , \half )}(\th_{ab}) =
\exp{(- \nu (\th _a \pauli ^{3} _a + \th _b \pauli ^{3} _b))}
S_{[a,b]}^{6V}(\th_{ab})
\exp{(\nu(\th _a \pauli ^{3} _a + \th _b \pauli ^{3} _b))}
\label{trans}
\ee
which is compatible with the YBE. Both the original and the
transformed \Sms~depend on the relative rapidity   $\th_{ab}= \th_a -
\th_b$ only, since they commute with the operator of the total isospin
projection $\h/2 = \pauli_a^3 + \pauli_b^3$.

The parameter $\nu$ may be chosen arbitrarily without violating both
the YBE and the unitarity.\footnote{ It should be real to preserve
real analyticity.} However, it must be related to the parameter $\g$
unambiguously once the crossing symmetry (\ref{cross}) is imposed.
Namely,
\be
\nu = 1 - {\g}/{\pi}.
\label{nu}
\ee
The transformation matrix from the eq.~(\ref{cross}) then reads
\be
U = \exp{(\imat \nu \pi (\pauli ^{3} _a - \pauli ^{3} _b)/2)}.
\label{cross-matr}
\ee
Note that the factor $Z(\th)$ must satisfy
two equations
\be
Z(\th) = Z(\ip - \th); ~~~~~~~{}
Z(\th) Z(-\th) =
\left(sinh{(\ig + \nt )}sinh{(\ig - \nt)}\right)^{-1}.
\label{factor}
\ee

After the parameter $\nu$ is chosen the \Sm~in question almost
coincides with that of the sine-Gordon model \cite{ZZ}. There are two
differences however. The first is due to the transformation
{}~(\ref{trans}) while the second is just the rescaling of the
rapidities. One should replace the parameter $\nu$ by $8 \pi/ \g_{SG}$
in order to relate the former to the latter exactly. The minimal
solution of the eqs.~(\ref{factor}) obtained in \cite{ZZ} reads
\be
Z(\th) = \frac{\imat}{\pi}
\G (1-\nu\al)\G (1-\nu(1-\al))
\prod_{n=1}^{\infty } R_n(\al) R_n(1-\al)
\label{Z1}
\ee
where $\al = \th/ \ip$ and
\be
R_n(\al) =
\frac
{\G (\nu(2n-\al))\G (1 + \nu(2n-\al))}
{\G (\nu(2n+1- \al))\G (1 + \nu(2n-1-\al))}
\label{Rn}
\ee
For $\nu >1$ this function has the well known breather poles in the
physical strip labeled by an integer $n$
\be
\al = 1 - n/\nu; ~~~~~~~~~\al = n/\nu~~~~~~~~~n \leq \nu.
\ee
The first series correspond to the zeros of $\sinh{(\ig - \nt)} =
\imat \sin{\nu\pi(1- \al)}$ while the second one appears in the cross
channel. When $\nu$ decreases and reaches the unity these poles leave
the physical strip. However, a new pole corresponding to the zero of
$\sinh{(\ig + \nt)} = \imat \sin{\nu \pi (1+ \al)}$ emerges for
$1/2<\nu <1$
\be
\al = 1/\nu -1, ~~~~~~~ or ~~~~~~~\th = \ig/\nu.
\label{pole1}
\ee
Due to the above inequality it is the only one in a physical strip.
This important fact will be discussed in the sect. \ref{fus}.
Meanwhile we are going to describe the symmetry of the solution.

\subsection[sym]{The symmetry.}

The quantum group ${\cal U}_q( sl(2))$ was dis\-co\-ve\-red by
Ku\-lish and Re\-she\-ti\-khin \cite{KR} as the symmetry of the above
solution and exploited to generalize it for the case of the higher
spins. We recall briefly the derivation following \cite{V}.

The YBE may be represented in the following form
\be
S_{[ab]}(\th_{ab})
T_{[0,ab]}(\th_{0a},\th_{0b})
=
T_{[0,ba]}(\th_{0b},\th_{0a})
S_{[ab]}(\th_{ab}) .
\label{ybe-TM}
\ee
where the monodromy-matrix
\be
T_{[0,ab]}(\th_{0a},\th_{0b}) =
S_{[0,a]}(\th_{0a}) S_{[0,b]}(\th_{0b}).
\label{TM}
\ee
is just the product of two \Sms.
Let us consider the YBE concentrating on the $\th_0$-dependence. Using
the eq.~(\ref{sixvert}) an explicit expression for the
monodromy-matrix easily derives of which three groups of terms (those
proportional to $e^{\pm 2\nt_0}$ and independent on $\th_0$
respectively) can be selected. Thus eq.~(\ref{ybe-TM}) splits into
three equations. Examining each of them separately and extracting the
terms proportional to $\pauli^{\pm}_0$ and to $\pauli^{3}_0$
respectively one can get the following linear equations for the \Sm
\bea
S_{[a,b]}(\th_a-\th_b) \Delta_{ab}(\pauli^{\lambda}) & = &
\Delta_{ba}(\pauli^{\pm}) S_{[a,b]}(\th_a-\th_b);
\label{interw} \\
S_{[a,b]}(\th_a-\th_b)  \delta_{ab}(\pauli^{\lambda}) & = &
\delta_{ba}(\pauli^{\lambda})
S_{[a,b]}(\th_a-\th_b),
\label{interw0}
\eea
where the operators attending the l.h.s of the
eq.~(\ref{interw}) are:
\bea
\Delta_{ab}(\pauli^{\pm}) \; = &\Delta(\pauli^{\pm}) &
= (\pauli^{\pm}_a \qsigm{b} + \qsig{a}
\pauli^{\pm}_b); \nonumber \\
\Delta_{ab}(\pauli^{3}) = &\Delta(\pauli^{3})&
= (\pauli^{3}_a + \pauli^{3}_b).
\label{coprod}
\eea
The operators in the r.h.s of the eq.~(\ref{interw}) differ from those
in the l.h.s by the permutation of $a$ and $b$:
\bea
\Delta_{ba}(\pauli^{\pm}) \; = &\Delb(\pauli^{\pm})&
= (\pauli^{\pm}_a \qsig{b} + \qsigm{a} \pauli^{\pm}_b); \nonumber \\
\Delta_{ba}(\pauli^{3}) = &\Delb(\pauli^{3})&
= (\pauli^{3}_a + \pauli^{3}_b)
\label{acoprod}
\eea
The operators involved in the eq.~(\ref{interw0}) are similar to the
above ones but depend on the rapidities as follows
\bea
\delta_{ab}(\pauli^{\pm}) \; = & \delta(\pauli^{\pm}) &=~
(\et{\mp 2}{a} \pauli^{\pm}_a \qsigm{b}
+ \et{\mp 2}{b} \qsig{a} \pauli^{\pm}_b)
\nonumber \\
\delta_{ab}(\pauli^{3}) \; = &\delta(\pauli^{3})& =~ (\pauli^{3}_a +
\pauli^{3}_b).
\label{t-coprd}
\eea
The expressions in the l.h.s.~(\ref{interw}) are related to the ones
in the r.h.s. by the permutation as in the eq.~(\ref{interw0})
\bea
\delta_{ba}(\pauli^{\pm}) \; = & \delb(\pauli^{\pm}) &=~
(\et{\mp 2}{a} \pauli^{\pm}_a \qsig{b}
+ \et{\mp 2}{b} \qsigm{a} \pauli^{\pm}_b)
\nonumber \\
\delta_{ab}(\pauli^{3}) = &\delta(\pauli^{3})& =~ (\pauli^{3}_a +
\pauli^{3}_b).
\label{t-acoprd}
\eea
This system turns out to be equivalent to the initial YBE~(\ref{ybe}).

It can be easily checked that each of the four triples of the
ope\-ra\-tors $\Delta_{ab}(\pauli^{\lambda})$,
$\Delta_{ba}(\pauli^{\lambda})$ and $\delta_{ab}(\pauli^{\lambda})$,
$\delta_{ba}(\pauli^{\lambda})$ provide four different (reducible)
matrix representations of the three elements $\e,\,\f,\,\h$ enjoying
the following commutation relations
\be
[ \e,\f] = \frac{\sin(\g \h)}{ \sin\g}; \;\;\;
 [ \h, \e] = 2 \e; \;\;\;  [ \h, \f] = - 2 \f
\label{Uq}
\ee
and generating an associative algebra called {\em quantum universal
enveloping algebra of} $sl(2)$ and denoted  ${\cal U}_q( sl(2)) \,
(q=e^{\imat \g})$. Below a shorthand notation ${\cal U}_q$ will be
used. When $\g \rightarrow 0$ the quantum universal enveloping algebra
degenerates into an ordinary enveloping algebra of $sl(2)$ denoted
${\cal U}( sl(2))$ or just ${\cal U}$ for brevity.

Note that the Pauli matrices with the usual commutation relations
\be
[ \pauli^{+}, \pauli^{-}] = \pauli^{3};\;\;\;\;\;
[\pauli^{3}, \pauli^{\pm}] = \pm \pauli^{\pm}
\label{pauli}
\ee
yield the simplest although trivial realization of~(\ref{Uq}) since
$\sin(\g \pauli^{3}) = \pauli^{3} \sin \g$.

Moreover, the six generators $\Delta_{ab}(\pauli^{\lambda})$,
$\Delta_{ba}(\pauli^{\lambda})$ or $\delta_{ab}(\pauli^{\lambda})$,
$\delta_{ba}(\pauli^{\lambda})$ combined provide a realization of the
$q$-deformed $A^{(1)}_1$ loop algebra referred as $\U_q^L$ below. The
latter is generated by six elements
$\e_1,\;\f_1,\;\h_1,\;\e_0,\;\f_0,\;\h_0$ obeying the following
commutation relations
\be
[ \e_i,\f_j] = \delta_{ij} \frac{\sin(\g \h_i)}{ \sin\g}; \;\;\;
 [ \h_i, \e_j] = 2 (-)^{i-j} \e_j; \;\;\;  [ \h_i, \f_j] = - 2 (-)^{i-j}
\f_j
\label{UAq}
\ee
just stating that the loop algebra includes two q-deformed $A_1$
subalgebras. These trivial relations are supplemented by Serre
relations
\bea
\e_i^3 \e_{j} - \e_{j} \e_i^3  - (1 + 2 \cos \g )\, (\e_i^2
\e_{j} \e_i -  \e_i \e_{j} \e_i^2) = 0;
\nonumber\\
\f_i^3 \f_{j} - \f_{j} \f_i^3  - (1 + 2 \cos \g )\, (\f_i^2
\f_{j} \f_i -  \f_i \f_{j} \f_i^2) = 0;
\label{serre}
\eea
for $i \neq j$

The correspondence between the generators $\Delta(\pauli_{\lambda}),
\; \delta(\pauli_{\lambda})$ is indicated in Table 1.

\bigskip
\begin{flushright}
{\sf Table 1}
\end{flushright}
\qquad
\begin{tabular}{|c|c|c|c|c|c|} \hline
 $ \Delta(\pauli^+) $& $\Delta(\pauli^-)$ & $ \Delta(\pauli^3) $ &
$\delta(\pauli^+) $ & $\delta(\pauli^-)$ & $- \delta(\pauli^3)$\\
\hline
$\e_1 $ & $\f_1 $ & $\h_1 $ & $\e_0 $ & $\f_0 $ & $ \h_0 $ \\
\hline
\end{tabular}

\bigskip

The above quantum algebra $\U_q$ is an one-parametric deformation of a
genuine $sl(2)$ enveloping algebra. From this view-point the mapping
$\Delta: {\cal U}_q \rightarrow End({\sf V}_a \otimes {\sf V}_b)$
called {\em co-product} and defined via~(\ref{coprod}) generalizes an
action of the $sl(2)$ for the case of two fundamental representations
while eq.~(\ref{acoprod}) yields the analogous mapping $\Delb$ for
${\sf V}_b \otimes {\sf V}_a$. The same mappings may be defined for
the direct product of any representations (even not necessarily
irreducible) and are in fact the algebra homomorphisms ${\cal U}_q
\rightarrow {\cal U}_q \otimes {\cal U}_q $:
\bea
\Delta_{12}(\h) ~\equiv &\Delta (\h)& = ~\h \otimes {\hat I} + {\hat
I} \otimes \h ; \\
\Delta_{12}(\e) ~\equiv &\Delta (\e)& = ~\e \otimes \qhm{} + \qh{}
\otimes \e ;\\
\Delta_{12}(\f) ~\equiv &\Delta (\f)& = ~\f \otimes \qhm{} + \qh{}
\otimes \f ; \label{copr}
\eea
\bea
\Delta_{21}(\h) ~\equiv &\Delb (\h)& = ~\h \otimes {\hat I} + {\hat I}
\otimes \h ;\\
\Delta_{21}(\e)  ~\equiv &\Delb (\e)& = ~\e \otimes \qhm{} + \qh{}
\otimes \e;\\
\Delta_{21} (\f)  ~\equiv &\Delb (\f)& = ~\f \otimes \qhm{} + \qh{}
\otimes \f ; \label{acopr}
\eea
This homomorphism may be extended to the whole $q$-deformed loop
algebra as follows
\bea
\Delta_{12}(\h_i) ~\equiv &\Delta (\h_i)& = ~\h_i \otimes {\hat I} +
{\hat I} \otimes \h_i ; \\
\Delta_{12}(\e_i) ~\equiv &\Delta (\e_i)& = ~\e_i \otimes \qhm{i} +
\qh{i} \otimes \e_i ;\\
\Delta_{12}(\f_i) ~\equiv &\Delta (\f_i)& = ~\f_i \otimes \qhm{i} +
\qh{i} \otimes \f_i ;
\label{Lcopr}
\eea
\bea
\Delta_{21}(\h_i) ~\equiv &\Delb (\h_i)& = ~\h_i \otimes {\hat I} +
{\hat I} \otimes \h_i ;\\
\Delta_{21}(\e_i)  ~\equiv &\Delb (\e_i)& = ~\e_i \otimes \qhm{i} +
\qh{i} \otimes \e_i;\\
\Delta_{21} (\f_i)  ~\equiv &\Delb (\f_i)& = ~\f_i \otimes \qhm{i} +
\qh{i} \otimes \f_i ;
\label{Lacopr}
\eea

The above co-multiplication combined with the following realization of
the $\U_q^L$ in terms of the Laurent polynomials in variable
$z=\et{2}{}$ over the underlying algebra $U_q$
\begin{flushright}
{\sf Table 2}
\end{flushright}
\qquad
\begin{tabular}{|c|c|c|c|c|c|} \hline
$\e $ & $\f $ & $\h $ & $z^{-1}\e $ & $z\f $ & $-\h $ \\
\hline
$\e_1 $ & $ \f_1 $ & $\h_1 $ & $\e_0 $ & $\f_0 $ & $\h_0 $ \\
\hline
\end{tabular}

\bigskip

may be used to generalize the operators entering the
eq.~(\ref{interw0})
\bea
\delta_{12}(\e) \; = &\delta (\e)& =\; \et{-2}{1} \e \otimes \qh{} +
\et{-2}{2} \qhm{} \otimes \e ;\\
\delta_{12}(\f) \;= &\delta (\f)& =\; \et{2}{1} \f \otimes \qh{} +
\et{2}{2} \qhm{} \otimes \f ;
\label{t-copr}
\eea
\bea
\delta_{21}(\e) \;= &\delb (\e)& =\; \et{-2}{1} \e \otimes \qhm{} +
\et{-2}{2} \qh{} \otimes \e;\\
\delta_{21}(\f) \;= &\delb (\f)& =\; \et{2}{1} \f \otimes \qhm{} +
\et{2}{2} \qh{} \otimes \f ;
\label{t-acopr}
\eea

Kulish and Reshetikhin~ \cite{KR} conjectured that the \Sm~
$S_{[a,b]}(\th_a-\th_b)$ solving the eqs.~(\ref{interw},\ref{interw0})
in ${\sf V}_a \otimes {\sf V}_b$ where ${\sf V}_a$ and ${\sf V}_b$ are
two arbitrary irreducible modules of ${\cal U}_q$ and Pauli matrices
should be substituted by the generators $\e, \; \f, \; \h$ provide the
solution of the YBE as well. This conjecture has been checked by
Jimbo~ \cite{J1} who proved the following

\begin{T}[Jimbo 1986]
Let
$\rho_a: {\cal U} \rightarrow End({\sf V_a}); \; (a=1,2,3)$
be three arbitrary finite-dimensional irreducible representations of
$sl(2)$. Assume that there exist three representations $\hat{\rho}_a$,
$\hat{\rho}_b$ and $\hat{\rho}_c$ of ${\cal U}_q$ such that
$\hat{\rho}_a \rightarrow \rho_a$ as $\g \rightarrow 0$.

\flushleft Then the linear equations
\bea
S_{[a,b]}(\th_a-\th_b) \Delta (\e) &=&
 \Delb (\e) S_{[a,b]}(\th_a-\th_b);
\label{int-e}\\
S_{[a,b]}(\th_a-\th_b) \delta (\f) &=&
\delb (\f) S_{[a,b]}(\th_a-\th_b)
\label{int-f0}
\eea
in ${End(\sf V}_a \otimes {\sf V}_b)$ has at most one solution for the
general value of $\g$.

If eqs.~(\ref{int-e}, \ref{int-f0}) admit a nontrivial solution
$S_{[a,b]}(\th_a-\th_b)$ it also satisfy the equations
\bea
S_{[a,b]}(\th_a-\th_b) \Delta (\f) &=&
 \Delb (\f) S_{[a,b]}(\th_a-\th_b);
\label{int-f}\\
S_{[a,b]}(\th_a-\th_b) \Delta (\h) &=&
 \Delb (\h) S_{[a,b]}(\th_a-\th_b);
\label{int-h}\\
S_{[a,b]}(\th_a-\th_b) \delta (\e) &=&
\delb (\e) S_{[a,b]}(\th_a-\th_b)
\label{int-e0}
\eea
Three solutions $S_{[a,b]}(\th_a-\th_b)$ for $(a,b) =
(1,2),\,(1,3),\,(2,3)$ satisfy the YBE in $End({\sf V}_1 \otimes {\sf
V}_2 \otimes {\sf V}_3)$.
\end{T}

\begin{rem}
An explicit solution for the above linear equations has been found in
\cite{KR} when one of the isospins either $J_a$ or $J_b$ is equal to
$\half$ (see the next subsection for a precise definition of the
isospin quantum numbers).
\bea
 S_{~[a,b]}^{(J,\half)}(\th) = Z(\th) \times
\qquad \qquad \qquad \qquad \qquad \qquad
\nonumber\\
\left\{ \sinh {\left( \nt  +  \imat \frac{\gamma}{2}
(1 + \h \otimes \pauli^3 \right)}
+ \imat \sin{\gamma} \left(e^{\nt}
\e \otimes \pauli^-
+  e^{- \nt}
\f \otimes \pauli^+
\right) \right\}
\label{half}
\eea
An examination of it shows that the above theorem is equivalent to the
following statement.

Let the isospins $J_a$ and $J_b$ be arbitrary and $J_0=\half$. Then
there is the only \Sm~$S^{(J_a,J_b)}_{[a,b]}(\th)$ satisfying the YBE
(\ref{ybe}). Once three such solutions $S^{(J_a,J_b)}_{[a,b]}(\th)$,
$S^{(J_a,J_c)}_{[a,c]}(\th)$ and $S^{(J_b,J_c)}_{[b,c]}(\th)$ for
arbitrary $J_a, J_b$ and $J_c$ do exist they satisfy the YBE
(\ref{ybe}).
\end{rem}

By means of this theorem an \Sm~$S_{[a,b]}(\th)$ for two arbitrary
irreducible representations ${\sf V}_a$ and ${\sf V}_b$ may be
constructed. This is a well known fusion process \cite{Kar,KRS} when
the \Sms~(\ref{sixvert}) are used as the elementary building blocks
for the \Sm~of the higher representations. We shall analyse this
procedure below considering it as a sort of bootstrap
equations~(\ref{boot})  and emphasizing the importance of the quantum
group invariance~(\ref{int-e},\ref{int-f0}). A few basic facts
concerning the finite-dimensional irreducible representations will be
important.

\subsection[irrep] {Description of the irreducible representations.}

The structure of the finite-dimensional irreducible representations of
${\cal U}_q$ for the generic values of $\g$ mainly repeats that of
${\cal U}$. They are labeled by the {\it highest weight} $J$ (the
maximal eigenvalue of $\frac{\h}{2}$) and will be denoted as $\{J\}$
below. The states inside each ${\cal U}_q$-module are labeled by the
{\it weight} $M$ (the eigenvalue of $\frac{\h}{2}$). The action of the
algebra $\U_q$ on the states is defined by the following formulas
\bea
\hh|J,M \rangle &= &M|J,M \rangle; \\
\e|J,M \rangle = d_M^J|J,M+1 \rangle ; &\;\;\;&
\f|J,M \rangle =  d_{M-1}^J|J,M -1\rangle ,
\label{matrep}
\eea
where
\be
d_M^J = \sqrt{[J+M+1]_q[J-M]_q};\;\;\;\;
[n]_q = \frac{\sin{\gamma n}}{\sin{\gamma}}.
\label{q-num}
\ee
Note that $\e|J,J \rangle = 0, \;\f|J,-J \rangle = 0$. The
normalization of the above basis is chosen in such a way that the
operators $\e$ and $\f$ are transposed to each other. There is a {\it
Casimir} operator
\be
\K = \f\e + \left[ \hh  \right]_q \left[ \hh + 1 \right]_q
\label{Casimir}
\ee
commuting with all generators $\e, \f, \h$ by virtue of the
commutation relations~(\ref{Uq}) and hence proportional to the
identity operator when restricted to an irreducible representation:
\be
\K|J,M \rangle = [J]_q [J+1]_q |J,M \rangle
\ee
\begin{rem}
For $J>\pi/\g$ the eigenvalues of Casimir may be negative indicating
that the states with a negative norm appear. The multiplets with the
higher isospin are thus forbidden in a physical sensible theory.
However, for sufficiently large $J$ the Casimir eigenvalues may become
positive again. The permitted J, though equidistant, may be separated
by the intervals larger then unity and this new group contains a
finite number of the spectral points. For generic $\g$ the permitted
spectrum consists of the infinite number of such groups known as the
Takahashi zones \cite{KiR,Tak,Frahm}. Moreover, this structure is
supplemented by another one including the so called odd states. All
these subtleties will ignored in the present paper and the
consideration will be restricted by the first Takahashi zone.
\end{rem}
The {\em Clebsch-Gordan} (C-G) decomposition
may be performed in two different ways.
\begin{enumerate}
\item
With respect to co-product $\Delta$:
\be
|J_a,M_a \rangle \otimes |J_b,M_b \rangle =
\sum_{J=|J_a-J_b|}^{|J_a+J_b|}
C_{M_a+M_b,M_a,M_b}^{~~~J,~~~~J_a,~J_b} |J,M_a+M_b\rangle
\label{CG}
\ee
The highest weight vector $|J,J \rangle$ in any irreducible
representation $\{J\}$ is annihilated by $\Delta (\e)$ while all other
states $|J,M \rangle$ are created by an operator $\left( \Delta
(\f)\right)^{J-M}$ acting on the highest weight vector.
\item
With respect to co-product $\Delb$.
\be
|J_a,M_a \rangle \otimes |J_b,M_b \rangle =
\sum_{J=|J_a-J_b|}^{|J_a+J_b|}
\overline {C}_{M_a+M_b,M_a,M_b}^{~~~J,~~~J_a,~J_b}
\overline{|J,M_a+M_b\rangle}
\label{aCG}
\ee
The structure of the C-G basis formed by the states
$\overline{|J,M\rangle}$ in this case is basically the same.
\end{enumerate}

We are going to discuss some representations of the whole $q$-deformed
loop algebra in what it follows. Since there is no central extension
the highest weight integrable representations do not exist. So, just
the formal Laurent polynomials over the irreducible representations
will be considered. They are certainly in one to one correspondence
with the usual finite dimensional irreducible representations of the
ordinary $q$-deformed $sl(2)$ algebra. However, as shown in the next
section, their C-G decomposition give rise to various representations
reducible with respect to the subalgebra $\U_q$.

\section[fus]{Massive fusion rules.}
\setcounter{equation}{0}

\subsection[+]{The irreducible bound states.}

Now suppose that two solutions of the YBE
$S_{[0,1]}^{(J_0,J_1)}(\th_{01})$ and
$S_{[0,2]}^{(J_0,J_2)}(\th_{02})$ are known (here the upper indices
denote the isospins of the irreducible representations 0,1,2). Let us
show how the \Sm~$S_{[0,12]}^{(J_0,J_1+J_2)}(\th)$ may be obtained.
Since both \Sms~must satisfy eqs.~(\ref{int-e},\ref{int-f0}) the
monodromy-matrix (\ref{TM}) also enjoys the following interwining
properties:
\bea
T_{[0,12]}(\th_{01},\th_{02}) \Delta_{012}(\e) &=&
\Delta_{120}(\e) T_{[0,12]}(\th_{01},\th_{02})
\label{tm-int-e}\\
T_{[0,12]}(\th_{01},\th_{02}) \delta_{012}(\f) &=&
\delta_{120}(\f) T_{[0,12]}(\th_{01},\th_{02})
\label{tm-int-f0}
\eea
where
\bea
\Delta_{012}(\e) & = & \left(\e \otimes \exh{-} +  \exp{(\imat
\frac{\g}{2} \h)} \otimes \Delta_{12}(\e) \right)
\label{Delta012}\\
\Delta_{120}(\e) & = & \left( \e \otimes \exh{} + \exp{(- \imat
\frac{\g}{2} \h)} \otimes \Delta_{12}(\e) \right)
\label{Delta120}
\eea
\bea
\delta_{012}(\f) & = & \left( \et{2}{0} \f \otimes \exh{}
 +  \qhm{} \otimes \delta_{12}(\f) \right)
\label{delta012}\\
\delta_{120}(\f)& = & \left( \et{2}{0} \f \otimes  \exh{-}
+ \qh{} \otimes \delta_{12}(\f) \right)
\label{delta120}
\eea
The monodromy-matrix~(\ref{TM}) acting in $\{J_0\} \otimes \{J_1\}
\otimes \{J_2\}$ performs a cyclic permutation in the co-product
$\Delta_{012}(\e)$ i.e. converts it into $\Delta_{120}(\e)$. Since the
co-product $\Delta_{12}(\e)$ is left intact by this operation it may
be transformed into the block-diagonal form by means of the C-G
decomposition in $\{J_1\}  \otimes \{J_2\}$. Multiplying both sides of
the eq.~(\ref{tm-int-e}) by the projectional operator $(1 \otimes
\Pro)$ one can verify that the operator \[ \left(1 \otimes \Pro)
T_{[0,12]}(\th_{01},\th_{02}) (1 \otimes \Pro \right) \]
enjoys the interwining property~(\ref{int-e}) in $\{J_0\} \otimes
\{J\}$ because the co-product operator $\Delta_{12}(\e)$ commutes with
the projector. This is not the case for the  operator
$\delta_{12}(\e)$ when the values of the rapidities $\th_1,\th_2$ take
generic values. These values may be adjusted, however, in such a way
that the the restriction of the operator $\delta(\f)$ to the
${J_1+J_2}$ module will coincide with that of $\Delta(\f)$.

\begin{prop}
Put $\nt_1 = -\imat \g J_2, \; \nt_2 = \imat \g J_1$. Then
\be
{\cal P}_{J_1, J_2}^{J_1+J_2} \delta_{12}(\f)
{\cal P}_{J_1, J_2}^{J_1+J_2} =
{\cal P}_{J_1, J_2}^{J_1+J_2}  \delta_{12}(\f)
{\cal P}_{J_1, J_2}^{J_1+J_2}.
\ee
\end{prop}

{\flushleft \underline{{\em Proof:}}
\quad First note that}
\be
\left[ \delta_{12}(\f), \Delta_{12}(\f) \right] =0.
\label{f-com-f}
\ee
Hence,
\be
 \delta_{12}(\f)  |J_1+J_2,M \rangle =
 \Delta_{12}(\f) |J_1+J_2,M \rangle
\label{d=D}
\ee
if and only if
\be
 \delta_{12}(\f)  |J_1+J_2, J_1+J_2 \rangle =
 \Delta_{12}(\f) |J_1+J_2, J_1+J_2 \rangle .
\label{in-idea}
\ee
The latter equality may be easily checked with the help of the
eqs.~(\ref{copr}) and (\ref{t-copr})  accounting for the fact that the
highest weight vector in the representation considered is just the
direct product of the two highest weight vectors:
\[  |J_1+J_2, J_1+J_2 \rangle  =
|J_1, J_1 \rangle   \otimes |J_1, J_1\rangle . \]

\begin{col}
The following bootstrap relation holds:
\be
{\cal P}_{J_1, J_2}^{J_1+J_2}
S_{[0,1]}^{(J_0,J_1)}(\th - \imat \gn J_2)
S_{[0,2]}^{(J_0,J_2)}(\th + \imat \gn J_1)
{\cal P}_{J_1, J_2}^{J_1+J_2} =
S_{[0,12]}^{(J_0,J_1+J_2)}(\th)
\label{fus}
\ee
\end{col}
In terms of parametrization~(\ref{mass-par}) the transition
$\{ J_1 \} \otimes \{ J_2 \} \rightarrow
\{ J_1+J_2 \} $ is characterized by the angles:
\be
\U ^{\J_1}_{\J_2,J_1+J_2} = \gn J_1; \; ~~~~
\U ^{\J_2}_{J_1+J_2,\J_1} = \gn J_2; \; ~~~~
\U ^{J_1+J_2}_{J_1,J_2} = \gn (J_1+J_2).
\label{angles}
\ee
Now a derivation of the mass spectrum for the particles with the
arbitrary isospin is straightforward:
\be
m_J = m_{\star}\sin{(\gn J)}
\label{mass}
\ee
where $m_{\star}$ is a common mass scale.

\subsection[lambda]{The analytic structure of the \Sm \\
for two irreducible multiplets.}

The multiplets above do not exhaust all the bound states in the
theory. The most direct way to see this is to examine an analytic
structure of the \Sm ~$ S_{[1,2]}^{(J_1,J_2)}(\th)$. We shall use a
modified version of analysis proposed by Jimbo~\cite{J3} for two
identical representations $(J_1=J_2)$ and going back to the work~
\cite{KRS}.

\begin{prop} \qquad
Let $\{J_1\}$ and $\{J_2\}$ be two  irreducible representations.
Then, the solution of the eqs.(~\ref{int-e},\ref{int-f0}) reads
\be
 S_{[1,2]}(\th) = \sum_{J=|J_1-J_2|}^{J_1+J_2}
\Lambda_J(\th) \Ppro
\label{Jsol}
\ee
where the operator
\be
\Ppro = \sum_{M=-J}^{J}
 \overline{|J,M\rangle} \langle J,M|.
\ee
Bra-vectors and ket-vectors belong to the C-G decompositions
(\ref{CG}) and (\ref{aCG}) respectively.
The "eigenvalues" $\Lambda_J(\th)$ are related to
each other recursively:
\be
\frac{\Lambda_{J}(\th)}{\Lambda_{J-1}(\th)} = -
\frac{\sinh{(\nt + \ig J)}}{\sinh{(\nt - \ig J)}}
\label{recurs}
\ee
\end{prop}

{\flushleft
\underline{{\em Proof:}} }
Due to eq.~(\ref{int-e},\ref{int-f},\ref{int-h}) the \Sm~enjoys the
interwining property with respect to the Casimir operators
\be
S_{[1,2]}(\th) \Delta(\K) = \Delb(\K) S_{[1,2]}(\th) .
\label{int-K}
\ee
This property combined with the analogous one for the operator $\h$
(\ref{int-h}) provides the selection rules for the \Sm~elements $
\overline{\langle J,M|}S_{[1,2]}(\th) |J^{\prime},M ^{\prime} \rangle$
which vanish if $J \neq J^{\prime}, M \neq M^{\prime}$.

The "eigenvalues"  $\Lambda(\th)$ may be extracted from the other
interwining equation (\ref{int-f0}) which is convenient to rewrite as
follows
\be
\Lambda_J(\th)  \langle J,M-1|\delta(\f)|J^{\prime}, M \rangle =
\overline{\langle J,M-1|} \delb(\f) \overline{|J^{\prime}, M \rangle}
\Lambda_{J^{\prime}}(\th)
\label{eig}
\ee
Though the number of the equations might seem to exceed the number of
eigenvalues, this is not the case because the most of the matrix
elements of $\delta(\f)$ just vanish and most of the non-vanishing are
related  to each other. Since the operator $\delta(\f)$ commutes with
$\Delta(\f)$ (see the eq. (\ref{f-com-f})) it always lowers the
isospin projection quantum number by one and may change the isospin
$J$ at most by one. Indeed, considering the matrix elements of the
product $\Delta(\f)$ and $\delta(\f)$ one can derive the following
relations
\be
d_M^{J^{\prime}}
\langle J,M-1|\delta(\f)|J^{\prime}, M \rangle
=
\langle J,M|\delta(\f)|J^{\prime}, M+1 \rangle
d_{M-1}^J
\label{f0-act}
\ee
where the matrix elements $d_M^J$ are given by the eq.~(\ref{q-num}).
Therefore any of the matrix elements may be expressed either through
the highest one \nobreak{
$\langle J,J^{\prime}-1|\delta(\f)|J^{\prime},J^{\prime}\rangle $}
or through the lowest one $\langle J,-J|\delta(\f)|J^{\prime},-J
\rangle$. On the other hand the highest matrix element does vanish for
all $J<J^{\prime}-1$ while the lowest one does vanish also for
$J>J^{\prime}-1$. Note that  the relations between the matrix elements
of the operator $\delb(\f)$ have exactly the same form as
(\ref{f0-act}). Thus the only independent equations are
\bea
\Lambda_{J \pm 1}(\th)
\langle J \pm 1,\mp J -1|\delta(\f)|J, \mp J \rangle
&=&
\overline{\langle J \pm 1, \mp J- 1|} \delb(\f)
\overline{|J, \mp J \rangle}
\Lambda_{J}(\th)
\nonumber\\
\langle J ,J-1|\delta(\f)|J,J \rangle &=&
\overline{\langle J ,J-1|} \delb(\f) \overline{|J, J \rangle}
\label{eig2}
\eea
Examining the $\th$-dependence of the above matrix elements one can
write them as
\[ \langle J \pm 1,\mp J -1|\delta(\f)|J, \mp J \rangle =
f^{\pm}_J(\g)
\exp{\left(\nu(\th_1 + \th_2)\right)}
\sinh{\left(\nt_{12} - \phi^{\pm}_J(\g)\right)}
\]
where $ f^{\pm}_J(\g)$ and $ \phi(\g)^{\pm}_J(\g)$ are some complex
functions of $\g$ which may be expressed through certain matrix
elements. Surprisingly enough $ \phi^{\pm}_J(\g)$ can be found
explicitly. It has been shown in \cite{J3} that
\be
\langle J \pm 1,\mp J -1|\delta(\f)|J, \mp J \rangle|_{\nt_{12} =
\mp \ig J} = 0
\label{break}
\ee
and consequently $ \phi^{\pm}_J(\g)= \mp \ig J $. Plugging these
matrix elements into the eq.~(\ref{eig2}) and accounting for the fact
that the substitution of  the operator $\delta(\f)$  by $\delb(\f)$
results just in the reversal of sign of $\g$ one can observe that the
relation (\ref{recurs}) holds up to the $\th$ independent factor. The
latter  is fixed by the unitarity (\ref{unit}) and the initial
condition (\ref{init}). Really, as it follows from the above arguments
the eigenvalues may be represented in the following form
\be
\Lambda_J(\th) = Z(\th,\g) \zeta_J(\g)
\prod_{L=J}^{J_1+J_2}  \sinh{(\nt + \imat \g L)}
\prod_{L=|J_1-J_2|+1}^{J-1}  \sinh{(\nt - \imat \g L)}
\label{exp-eig}
\ee
where the common factor $Z(\th,\g)$ and the coefficients $\zeta_J(\g)$
are subject to the following constraints
\[
Z(\th,\g) Z(-\th,\g)\left(\zeta_J(\g) \right)^2 =
\prod_{L=|J_1-J_2|+1}^{J_1+J_2}
cosech{(\nt + \imat \g L)} cosech{(\nt - \imat \g L)}; \nonumber
\]
\[
Z(\th=0,\g) \zeta_J(\g) = (-1) ^{J_1+J_2-J} . \nonumber
\]
These constraints can be easily splitted giving
\be
Z(\th,\g) Z(-\th,\g) =
 \prod_{L=|J_1-J_2|}^{J_1+J_2}
cosech{(\nt + \imat \g L)} cosech{(\nt - \imat \g L)} ; \nonumber \\
\ee
\be
Z(\th=0,\g) = 1; ~~~~~~~~~~~~ \zeta_J(\g) = (-1) ^{J_1+J_2-J}
\label{Z-factor}
\ee
The validity of the relations (\ref{eig2}) now follows from the
Jimbo's theorem and the explicit construction of the \Sm~(\ref{fus}).

Thus, the proof is  completed.

\medskip
The relation (\ref{Z-factor}) is not sufficient to calculate a
function $Z(\th)$ and in this way to find out the poles of the \Sm~in
the physical strip corresponding to some bound states. Unfortunately,
I have not found any straightforward arguments in favor of the
crossing-symmetry of of the \Sm~(\ref{fus}) for the generic values of
the isospins $J_1$ and $J_2$. However, there are several solutions
known explicitly. Those with $J_1 = \half$ and $J_2$ arbitrary (or
vice versa) were found in \cite{KR}. Another one with $J_1 = J_2 = 1$
was computed in \cite{FZ}. The above examples are crossing-symmetric.
So, the following conjecture seems to be reasonable.
\begin{con}
The \Sm~(\ref{fus}) for the generic values of the isospins
$J_1$ and $J_2$ possess the crossing-symmetry (\ref{cross}) with the
unitary matrix
\be
U = \exp{(\imat \nu \pi (\h \otimes {\hat I} - {\hat I} \otimes \h
)/2)}.
\label{cross-matr-gen}
\ee
This crossing-symmetry leads to the equation for the Z-factor
\be
Z(\th) = Z(\ip - \th).
\label{Z-cross}
\ee
\end{con}
\medskip

The minimal solution of the eqs.~(\ref{Z-factor},\ref{Z-cross}) having
the poles at all the points $\th = \ig J/\nu;~(|J_1-J_2| \leq J \leq
J_1+J_2))$ reads
\be
Z(\th) = \prod_{J=|J_1 - J_2|}^{J_1+J_2} Z_J(\th)
\ee
where
\be
Z_J(\th) = \frac{\imat}{\pi}
\G (J\beta-\nu\al)\G (J\beta-\nu(1-\al))
\prod_{n=1}^{\infty } R_n^J(\al) R_n^J(1-\al).
\label{ZJ}
\ee
Here $\beta=\g/\pi=1-\nu$ and
\be
R_n^J(\al) =
\frac
{\G (1-J\beta-\nu(2(n-1)+\al))\G(J\beta - \nu(2n+\al))}
{\G (1-J\beta-\nu(2n-1+ \al))\G (J\beta- \nu(2n-1+\al))}.
\label{RnJ}
\ee
While the function $R_n^J(\al)$ has no poles in the physical strip $0
< \al < 1$ the pre-factor $\G$-functions do have them at
\be
\begin{array}{lcl}
\al = \beta J/\nu;&~~~&\th =\ig J/\nu;\\
\al = 1-\beta J/\nu;& &\th =\ip - \ig J/\nu;
\end{array}
\ee
provided $\g J/\nu<\pi$.

The value of $\nt= \ig (J_1+J_2)$ at which the bound states
$|J_1+J_2,M \rangle$ arise is just the value when the ratio
(\ref{recurs}) for $J=J_1+J_2$ vanishes. The tensor representation
$\{J_1\} \otimes \{J_2\}$ becomes reducible with respect to the total
algebra $U_q^L$. This happens because, the usual mapping $\delb(\f):
\{J_1+J_2\} \rightarrow \{J_1+J_2\} \oplus \{J_1+J_2-1\}$ due to the
eq. (\ref{break}) degenerates into $\delb(\f): \{J_1+J_2\} \rightarrow
\{J_1+J_2\}$ making the module $\{J_1+J_2\}$ irreducible.

This is not the only value of $\th$ for which the degeneracy occurs.
Consider e.g. $\nt= \ip-\ig(|J_1-J_2| +1)$. (Here $\ip$ is added in
order to choose $\th$ within the physical strip $0 < \th < \pi$.) This
value corresponds to the decoupling of the lowest representation
$\{|J_1-J_2|\}$ of the C-G series (\ref{CG}). Inserting $\nu$ from
(\ref{nu}) one easily finds the Euclidean angle
\be
\U ^{|J_1-J_2|}_{J_1,J_2} = \ip - \gn |J_1-J_2|.
\label{min-angle}
\ee
Substituting it into the eq.(\ref{add}) and using the momentum
conservation (\ref{cons:e-m}) one obtains two other angles
characterizing the reaction
\be
\begin{array}{lclclcl}
\U ^{\J_1}_{\J_2,|J_1-J_2|} =  \gn J_2; &
&\U ^{\J_2}_{|J_1-J_2|,\J_1} = \pi-\gn J_1;& & for& &J_1>J_2;
\nonumber\\
\U ^{\J_1}_{\J_2,|J_1-J_2|} = \pi-\gn J_2; &
&\U ^{\J_2}_{|J_1-J_2|,\J_1} = \gn J_1;& & for& &J_1<J_2;
\nonumber\\
\U ^{\J_1}_{\J_2,|J_1-J_2|} = \half \pi; &
&\U ^{\J_2}_{|J_1-J_2|,\J_1} = \half \pi\;&  &for& & J_1=J_2;
\end{array}
\label{dif-angles}
\ee
Plugging the above values into the expressions (\ref{mass-par}) for
the masses of the particles participating in the reaction yields
\be
m_{J_1} = m_{\star} \sin{\gn J_1}; ~~~~
m_{J_2} = m_{\star}\sin{\gn J_2};~~~~
m_{|J_1-J_2|} = m_{\star}\sin{\gn |J_1-J_2|}
\ee
in full agreement with the spectrum (\ref{mass}).\footnote{Notice that
the iso-scalar particles are massless.}

This is not an end of the story however. A similar considerations
apply to all the values of $\th$ for which any ratio (\ref{recurs})
either vanishes or turns into infinity.

In commodity with (\ref{break}) when the rapidity takes the
corresponding value $\th = \ig J/\nu$ the operator $\delta(\f)$ maps
the $\U_q$-module $\{J\} \rightarrow \{J\} \oplus \{J+1\}$ while the
usual mapping to $\{J-1\}$ disappears. Hence, an irreducible module $
\{J_1\} \otimes \{J_2\}$ becomes reducible with respect to the whole
algebra $U_q^L$ and the following representation decouples:
\be
\{J_1, J_2 \|J_{+}\} =
\{J\} \oplus \{J+1\} \oplus \cdots \oplus \{J_1+J_2\}
\label{reducible+}
\ee
There  exits another series of the rapidities $\th = \imat(\pi - \g
J)/\nu$ for which the module $ \{J_1\} \otimes \{J_2\}$ splits since
the mapping $\{J\} \rightarrow \{J-1\} \oplus \{J\} \oplus \{J+1\}$
degenerates into
$\{J\} \rightarrow \{J-1\}\oplus  \{J\} $. The decoupling
representation is now
\be
\{J_1, J_2 \|J_{-}\} =
\{J\} \oplus \{J-1\} \oplus \cdots \oplus \{|J_1-J_2|\}
\label{reducible-}
\ee
The above super-multiplets labeled by three quantum numbers $J_1,J_2$
and $J_{\pm}$, contain new bound states having equal masses:
\be
m_{\{J_1,J_2\|J_{\pm}\}}^2 = m_{J_1}^2 + m_{J_2}^2
\pm 2 m_{J_1} m_{J_1} \cos{(\gn J)}
\label{mass+-}
\ee
The  scattering angles $U^{\J_1}_{\J_2,J_{\pm}} $ and
$\U^{\J_2}_{J_{\pm},\J_1}$ should be defined through the following
transcendental equations:
\bea
\sin{\g J_1} \sin{\U^{\J_1}_{\J_2,J_{\pm}}}& = &
\sin{\g J_2} \sin{\U^{\J_2}_{J_{\pm},\J_1}} \nonumber\\
\U^{\J_1}_{\J_2,J_{+}} + \U^{\J_2}_{J_{+},\J_1} = \gn J; & ~ &
\U^{\J_1}_{\J_2,J_{-}} + \U^{\J_2}_{J_{-},\J_1} = \pi - \gn J .
\label{angles+-}
\eea

\medskip

Now we are in position to prove that the multiplets under
consideration really emerge from bootstrap equations. The following
generalization of the fusion procedure proposed by Karowsky \cite{Kar}
for the \Sms ~being rational functions of rapidities gives the unitary
\Sm~satisfying the YBE for the representation considered.

\begin{prop}
\sloppy{
Define the following operators projecting onto the sub\-spaces
(\ref{reducible+},\ref{reducible-})}
\bea
{\cal Q}_{\{J_1,J_2\|J_-\}}
 = \sum_{L=|J_1-J_2|}^{J}
\mu_L^{J_-} \pro; & &
{\cal Q}_{\{J_1,J_2\|J_-\}}^{-1}
 = \sum_{L=|J_1-J_2|}^{J}
\left( \mu_L^{J_-} \right)^{-1} \pro; \quad
\nonumber \\
{\cal Q}_{\{J_1,J_2\|J_+\}}
 =  \sum_{L=J}^{J_1+J_2} \mu_L^{J_+} \pro; &&
{\cal Q}_{\{J_1,J_2\|J_+\}}^{-1}
 = \sum_{L=J}^{J_1+J_2}
\left( \mu_L^{J_+} \right)^{-1} \pro; \quad
\label{j-projectors}
\eea
where
\be
\left. \left(\mu_L^{J_+}\right)^2
 =  \frac{\Lambda_L(\th)}{Z(\th)} \right|_{\th = \frac{\ig J}{\nu}};
\quad
\left. \left(\mu_L^{J_-}\right)^2
= \frac{\Lambda_L(\th)}{Z(\th)}\right|_{\th =\imat\pi- \frac{\ig
J}{\nu}}
\label{mu}
\ee
and the reciprocality may be understood literally once a restriction
to the subspace $\{J_1,J_2\|J_{\pm}\}$ is imposed. Then the unitary
\Sm~solving the YBE and respecting the condition of the real
analyticity reads
\bea
S_{[0,12]}^{(J_0,J_{\pm})}(\th) = \qquad \qquad
\qquad \qquad \qquad \qquad \qquad \qquad \qquad \qquad
\nonumber\\
{\cal Q}_{\{J_1,J_2\|J_{\pm}\}}^{-1}
S_{[0,1]}^{(J_0,J_1)}(\th - \imat\U^{\J_1}_{\J_2,J_{\pm}})
S_{[0,2]}^{(J_0,J_2)}(\th + \imat \U^{\J_2}_{J_{\pm},\J_1})
{\cal Q}_{\{J_1,J_2\|J_{\pm}\}}
\label{hard-fus}
\eea
\end{prop}
The proof is somewhat cumbersome and is presented in Appendix. Here I
would like to emphasize that the choice of the eigenvalues for the
operators (\ref{j-projectors}) is determined by the condition of real
analyticity and, hence, by unitarity while it is irrelevant for the
YBE.

\bigskip

The spectral decomposition analogous to (\ref{Jsol}) for the \Sm
{}~involving one of the representations $\{J_1,J_2\|J_{\pm}\}$ and any
irreducible one just as that involving two reducible representations
hasn't been yet computed. Nevertheless, there is no doubt that the new
bound states corresponding to highly reducible representations do
exist. The reduction of these super-multiplets should split them into
a finite number of the irreducible ones with the isospins $J=J_{min},
J_{min}+1, \ldots J_{max}-1, J_{max}$. The multiplicities of these
constituents should be equal to one either for the maximal or for the
minimal isospins and ought to be greater then one for the other
species. Their structure will be investigated elsewhere. However, I
would like to finish this section with a following conjecture.
\begin{con}
The isotopic structure of multiplets together with the corresponding
scattering angles may be found recursively. To do this one should
consider all possible fusions of the multiplets already found. Any
direct product of two multiplets would be irreducible with respect to
$\U_q^L$ for the generic values of the rapidities. There exist the
imaginary value of the relative rapidity for which it becomes
reducible. This value gives the scattering angles while extracting the
irreducible part of the direct product one gets the isotopic contents
of the new multiplet.
\end{con}

\section{Concluding remarks}

There are two main results obtained in the present work. First, a lot
of new bound states yet unknown in the context of the sine-Gordon
model have been found. Second, the particle contents of this system is
shown to be in one to one correspondence with a set of very peculiar
representations of the quantum loop algebra. An algorithm is proposed
for generating these representations recursively.

In my opinion, it is tempting to match this \PEST~ to some integrable
perturbation of the WZNW \CFT. The main reasons why this
correspondence seems to be plausible is that the \Sm is invariant
under the action of the $q$-deformed loop algebra. Though this algebra
does not have an anomalous central extention the latter  might be
produced dynamically. In fact the reqirements of the unitarity and
crossing permit only a finite number of the irreducible multiplets
equal to the integer part of $\pi\nu/\g$ just as it happens in the
WZNW model with the Kac-Moody central charge $k=[\pi\nu/\g]$. The
reducible representations might be interpreted as some combinations of
the descendent states. Since the central charge is known to serve as a
measure for the number of degrees of freedom of the system \cite{Zrg}
and the number of irreducible representations decreases with the
increase of $\g$ it looks like that this parameter describes the
Zamolodchikov's \cite{Zrg} renormalization group flow along one of the
trajectories. The rational values of $\g$ where some of the states
become massless might correspond to some critical points.

On the other hand the integrable deformations of the minimal models
are also known to have the symmetry algebra coinciding with the
$q$-deformed $A_1^{(1)}$ algebra \cite{ResSm,BerCl,Mathur}. So, to
establish the exact correspondence between the \PEST ~in question and
some perturbed \CFT ~it is necessary to calculate at least the central
charge of the system.

The most direct way to compute the universal characteristics of the
model - its central charge and the primary conformal dimensions is the
finite size corrections approach \cite{FSC,fscc,fsAf} modified
specifically for the \PEST ~by Al.~Zamolodchikov \cite{AZ}.  A
thermodynamic version of the \BA ~applies to a massive integrable
quantum field theory at a temperature much lower than the smallest of
the masses but sufficiently high to treat the system thermodynamically
allowing to calculate a specific heat proportional to the central
charge \cite{fscc,fsAf}. Moreover, examining the collective
temperature excitations with the complex rapidities one can in
principle obtain the conformal dimensions \cite{M}.

A sort of such calculation has been already done in \cite{KiR} and the
result is consistent with the WZNW interpretation. However,  only the
irreducible multiplets were taken into account in the cited work. The
derivation of the Bethe ansatz for the infinite system of excitations
looks as a tremendous task at present accounting for the fact
especially that the above excitations have not been described yet and
the corresponding \Sm~ has not been calculated.

Investigating the structure of the bound states one encounters the
following mathematical problem. The realization of the algebra
$\U_q^L$ given in Table 2 is the only one for which an algebraic
definition for the operators $\e_0, \f_0$ in terms of the $\e, \f$ is
known. This realization is naturally associated with the irreducible
representations of the $q$-deformed $sl(2)$ algebra. The action of the
operators $\e_0$ and $\f_0$ on the direct product of the irreducible
representations is described by the eq. (\ref{t-copr}) quite
satisfactory but the same action on the states of the module $\{J_1,
J_2\|J\}$ should be determined recursively.

Finally, the relation of the model in question to the genuine
sine-Gordon model is far from being clear. Their $Z$-factors for the
\Sms~of the fundamental representations are in fact different
functions of $\th$ though they do coincide at $\nu=1$ or
$\g_{SG}=8\pi$. At this point the \Sm~ degenerates into the identity
operator since the reflection amplitude of the soliton-antisoliton
scattering vanishes while the transition amplitude becomes equal to
the amplitude of scattering of the two identical particles and both
equal to unity. Thus, in my view this point might be the point of
singularity for the \PEST~ considered i.e. the boundary point of two
different regimes. Unfortunately, neither the semiclassical
approximation nor perturbative approach can't be applied for the
values of coupling constants close to unity. So, a similarity of the
\Sms~may occur just formal. However, if the physical significance of
the relation would be established it could provide an interesting
example of duality between the weak- and strong-coupling regimes of
the sine-Gordon model.

\pagebreak[4]

In any case the system seems to me deserving further attention due to
its highly non-trivial structure. An examination of the latter might
provide new profound insights in the near-critical universality.

\medskip

The same algebraic methods are applicable for the systems with the
higher underlying symmetries. In particular the solutions of the YBE
involving two arbitrary irreducible representations (not necessarily
rectangular as in \cite{OW,RW}) can be found. But this is a subject of
the future publication.

\bigskip

\underline{Acknowledgments:} I am grateful to H.~Braden for several
very stimulating discussions and to A.~Zamolodchikov for the valuable
remarks. I am indebted to Prof. Howard J. Schnitzer for his support. I
would like to thank Princeton University where this work was finished
and especially A.~Polyakov for the hospitality.

\setcounter{equation}{0}
\renewcommand{\theequation}{A.\arabic{equation}}
\section*{Appendix}
{\underline{\em Proof}} of the Proposition 4.

It would be convenient to use the notion of the monodromy-matrix
(\ref{TM}) introduced previously for the product of two \Sms ~just as
those entering the eq. (\ref{hard-fus}). The monodromy-matrices
themselves obey a sort of the YBE:
\bea
S_{[ab]}(\th_{ab})
T_{[a,12]}(\th_{a1},\th_{a2})
T_{[b,12]}(\th_{b1},\th_{b2})
=~~~~~~~~~~~~~~~~~~~~~~~~~~\nonumber\\
{}~~~~~~~~~~~~~~~~~~~~~~~~~~~
T_{[b,12]}(\th_{b1},\th_{b2})
T_{[a,12]}(\th_{a1},\th_{a2})
S_{[ab]}(\th_{ab})
\label{TT-ybe}
\eea
So, the \Sms~(\ref{hard-fus}) could enjoy the YBE if the projectional
operators (\ref{j-projectors}) did not interfere. Indeed, the eq.
(\ref{TT-ybe}) may be sandwiched between the operators ${\cal
Q}_{\{J_1,J_2\|J_{\pm}\}}^{-1}$ and ${\cal Q}_{\{J_1,J_2\|J_{\pm}\}}$,
so it's sufficient to show that
\bea
{\cal Q}_{\{J_1,J_2\|J_{\pm}\}}^{-1}
T_{[a,12]}(\th_{a1},\th_{a2})
T_{[b,12]}(\th_{b1},\th_{b2})
{\cal Q}_{\{J_1,J_2\|J_{\pm}\}} = \nonumber\\
\sum_{L}
{\cal Q}_{\{J_1,J_2\|J_{\pm}\}}^{-1}
T_{[a,12]}(\th_{a1},\th_{a2})
\pro
T_{[b,12]}(\th_{b1},\th_{b2})
{\cal Q}_{\{J_1,J_2\|J_{\pm}\}}
\label{restrict-ybe}
\eea
provided
$ \th_{a1}-\th_{a2}=\th_{b1}-\th_{b2}=\ig J/\nu $
or $\imat (\pi-\g J)/\nu $ .
The summation runs over $J \leq L \leq J_1+J_2$ or $|J_1-J_2|\leq L
\leq J$ respectively.

Substituting the decomposition (\ref{Jsol}) for
$S_{[12]}(\th)$ into the YBE
(\ref{ybe-TM}) involving two irreducible representations
$\{J_1\}$ and $\{J_2\}$
one may represent them in the following form
\bea
\Lambda_J(\th)  \langle J,M|
T_{[0,12]}(\th_{01},\th_{02})
|J^{\prime}, M ^{\prime} \rangle =
{}~~~~~~~~~~~~~~~~~~~~~~~~~~~~~~~~~~~~~~\nonumber\\
\qquad
\overline{\langle J,M|}
T_{[0,21]}(\th_{02},\th_{01})
\overline{|J^{\prime}, M ^{\prime} \rangle}
\Lambda_{J^{\prime}}(\th)
\label{ybe-eig}
\eea
The subsequent application of the eqs.~(\ref{ybe-TM},\ref{ybe-eig}) to
the product of two mo\-no\-dro\-my-matrices leads to the following
identities
\bea
\qquad \Lambda_J(\th)  \langle J,M|
T_{[a,12]}(\th_{a1},\th_{a2})
T_{[b,12]}(\th_{b1},\th_{b2})
|J^{\prime}, M^{\prime} \rangle =~~~~~~~~~~~~{}
\nonumber\\
\sum_{J^{\prime\prime}=|J_1-J_2|}^{J_1+J_2}
\sum_{M^{\prime\prime} = -J^{\prime\prime}}^{J^{\prime\prime}}
\langle J,M|
T_{[a,12]}(\th_{a1},\th_{a2})
|J^{\prime\prime}, M^{\prime\prime} \rangle
\times~~~~~~~~~~~~~~~~~~~~~~~~~~~~{}
\nonumber\\~~~~~~~~ {}
\Lambda_{J^{\prime\prime}}(\th)
\overline{\langle J^{\prime\prime},M^{\prime\prime}|}
T_{[b,21]}(\th_{b2},\th_{b1})
\overline{|J^{\prime}, M^{\prime} \rangle} =
\nonumber\\
\overline{\langle J,M|}
T_{[a,21]}(\th_{a2},\th_{a1})
T_{[b,21]}(\th_{b2},\th_{b1})
\overline{|J^{\prime}, M^{\prime} \rangle}
\Lambda_{J^{\prime}}(\th) ~~~~~~~~~~~~~{}
\label{identity-eig}
\eea
Putting $\th_{12}=\ig J/\nu$ and using the YBE (\ref{ybe-eig}) one can
easily derive the following relations
\bea
\langle J^{\prime},M^{\prime}|
T_{[a,12]}(\th_{a1},\th_{a2})
T_{[b,12]}(\th_{b1},\th_{b2})
|J^{\prime\prime}, M^{\prime\prime} \rangle
=~~~~~~~~~~~~~~~~~~~~~~~~~~~~~~~{}
\nonumber\\
\sum_{L=J}^{J_1+J_2}
\langle J^{\prime},M^{\prime}|
T_{[a,12]}(\th_{a1},\th_{a2})
|J^{\prime}, M^{\prime} \rangle \times
\nonumber\\
\overline{\langle J^{\prime},M^{\prime}|}
T_{[b,21]}(\th_{b2},\th_{b1})
\overline{|J^{\prime\prime}, M^{\prime\prime} \rangle}
 =~~~~~~~~~~~~~~~~~~~~~~~~~~~~~~~{}
\nonumber\\
\sum_{L=J}^{J_1+J_2}
\overline{\langle J^{\prime},M^{\prime}|}
T_{[a,21]}(\th_{a2},\th_{a1})
\pro
T_{[b,21]}(\th_{b2},\th_{b1})
\overline{|J^{\prime\prime}, M^{\prime\prime} \rangle} \qquad
\label{restrict}
\eea
provided both the quantum numbers $J^{\prime}$ and $J^{\prime\prime}$
labeling the matrix elements are restricted by an inequality $J \leq
J^{\prime} , J^{\prime\prime} \leq J_1+J_2$. Note that the summation
ranges over the irreducible modules $\{L\} \i \{J_1,J_2\|J_+\}$ . Thus
the above relations almost coincide with the eq. (\ref{restrict-ybe})
we are going to proof. The latter may be reduced to the former by a
similarity transformation of a monodromy-matrix with the matrices
(\ref{j-projectors}) \footnote{This is a similarity  transformation
only in the irreducible module $\{J_1,J_2\|J_+\}$.}
\be
T_{[a,12]}(\th_{a1},\th_{a2}) \longrightarrow
{\cal Q}_{\{J_1,J_2\|J_+\}}^{-1}
T_{[a,12]}(\th_{a1},\th_{a2})
{\cal Q}_{\{J_1,J_2\|J_+\}}
\label{similar}
\ee
and the same for
$T_{[b,12]}(\th_{b1},\th_{b2})$.
The similarity transformation does not certainly affect the validity
of the YBE, but on the other hand it's not clear yet what it is needed
for. The explanation is rather simple - the transformation
(\ref{similar} is necessary due the unitarity requirement.

\smallskip
Recall that the eq. (\ref{unit}) stems from the YBE and the initial
conditions (\ref{init}) only. However, this is not sufficient for the
unitarity of the \Sm. The real analyticity (\ref{analyt}) must hold in
addition
\be
\left(S_{[0,12]}(\th)\right)^{\dagger} =
S_{[12,0]}(-\th^{\ast})
\ee
In terms of the matrix elements of the \Sm~(\ref{hard-fus}) the above
equation reads
\bea
\mu_{J^{\prime}}^{J_{\pm}}
\left(\mu_{J^{\prime\prime}}^{J_{\pm}}\right)^{-1}
{\hat {\cal T}}_0
\langle J^{\prime\prime},M^{\prime\prime}|
S_{[0,1]}(\th - \imat\U^{\J_1}_{\J_2,J_{\pm}})
S_{[0,2]}(\th + \imat \U^{\J_2}_{J_{\pm},\J_1})
|J^{\prime}, M^{\prime} \rangle^{\ast}
= \nonumber\\
\left( \mu_{J^{\prime}}^{J_{\pm}} \right)^{-1}
\mu_{J^{\prime\prime}}^{J_{\pm}}
\langle J^{\prime},M^{\prime}|
S_{[1,0]}(-\th^{\ast} - \imat\U^{\J_1}_{\J_2,J_{\pm}})
S_{[2,0]}(-\th^{\ast} + \imat \U^{\J_2}_{J_{\pm},\J_1})
|J^{\prime\prime}, M^{\prime\prime} \rangle
\nonumber
\eea
where $ {\cal T}_0$ is the operator of transposition in the 0-th
space.

Since the complex conjugation converts  the C-G decomposition
(\ref{CG}) into (\ref{aCG}) at real $\g$  the above equation may be
rewritten as
\bea
\left( \mu_{J^{\prime}}^{J_{\pm}} \right)^2
\overline{
\langle J^{\prime\prime},M^{\prime\prime}|}
{\hat {\cal T}}_0
S_{[0,2]}^{\ast} (\th + \imat \U^{\J_2}_{J_{\pm},\J_1})
{\hat {\cal T}}_0
S_{[0,1]}^{\ast} (\th - \imat\U^{\J_1}_{\J_2,J_{\pm}})
\overline{
|J^{\prime}, M^{\prime} \rangle}
= \nonumber\\
\left(\mu_{J^{\prime\prime}}^{J_{\pm}}\right)^2
\langle J^{\prime},M^{\prime}|
S_{[1,0]}(-\th^{\ast} - \imat\U^{\J_1}_{\J_2,J_{\pm}})
S_{[2,0]}(-\th^{\ast} + \imat \U^{\J_2}_{J_{\pm},\J_1})
|J^{\prime\prime}, M^{\prime\prime} \rangle
\nonumber
\eea
where each of the transposition operators in the l.h.s. of the last
equality acts on the nearest \Sm~only. Finally, using the property of
real analyticty for any of the \Sms~inside the matrix elements one
comes to the equality
\bea
\left( \mu_{J^{\prime}}^{J_{\pm}} \right)^2
\overline{
\langle J^{\prime\prime},M^{\prime\prime}|}
S_{[0,2]}(-\th^{\ast}  + \imat \U^{\J_2}_{J_{\pm},\J_1})
S_{[0,1]}(-\th^{\ast}  - \imat\U^{\J_1}_{\J_2,J_{\pm}})
\overline{
|J^{\prime}, M^{\prime} \rangle}
= \nonumber\\
\left(\mu_{J^{\prime\prime}}^{J_{\pm}}\right)^2
\langle J^{\prime},M^{\prime}|
S_{[1,0]}(-\th^{\ast} - \imat\U^{\J_1}_{\J_2,J_{\pm}})
S_{[2,0]}(-\th^{\ast} + \imat \U^{\J_2}_{J_{\pm},\J_1})
|J^{\prime\prime}, M^{\prime\prime} \rangle
\nonumber
\eea
which due to the eq. (\ref{mu})
is just the particular case of the eq. (\ref{ybe-eig}).

\smallskip

\flushleft{The proof is completed.}

\end{document}